\documentclass[11pt]{article}
 \usepackage{cite}
 \usepackage{times}
 \usepackage{graphicx}
 \usepackage{hyperref}
 \usepackage{multicol}
 \usepackage[text={181.5mm,243mm},centering,a4paper]{geometry}
 \usepackage[labelfont={sf,bf,small},textfont={sf,small}]{caption}
 \usepackage{color}
 \usepackage{fancyhdr}
\definecolor{gray}{rgb}{0,0,0.6}
\definecolor{dgreen}{rgb}{0,0.5,0}

\makeatletter
\def\@cite#1#2{$^{\mbox{\scriptsize #1\if@tempswa , #2\fi}}$}
\renewcommand\@biblabel[1]{#1.}
\makeatother

\newcommand{\fig}[1]{Fig.\,\ref{#1}}
\newcommand{\figs}[1]{Figs.\,\ref{#1}}
\newcommand{\equ}[1]{Eq.\,(\ref{#1})}

\graphicspath{{./}{figs/}}
\newcommand{\aap}{Astron.\ \& Astrophys.}
\newcommand{\apj}{Astrophys. J.}
\newcommand{\apjl}{Astrophys. J. Letters}
\newcommand{\jgr}{J. Geophys. Res.}
\newcommand{\solphys}{Sol. Phys.}
\newcommand{\pasj}{Publ. Astron. Soc. Japan}

\begin{document}

\newgeometry{text={181.5mm,243mm},centering}
\setlength{\columnsep}{6mm}

\captionsetup[figure]{labelfont={sf,bf,small},textfont={sf,footnotesize}}

% TITLE %%%%%%%%%%%%%%%%%%%%%%%%%%%%%%%%%%%%%%%%%%%%%%%%%%%%%%%%%%%%%%%%%%%%%%%

\noindent
{\Large\sf\bfseries 
Magnetic Jam in the Corona of the Sun
}
\\[2ex]
{
F.~Chen,$^{1}$ 
H.~Peter,$^{1}$ 
S.~Bingert,$^{2}$ 
M.C.M.~Cheung$^{3}$
}
\\[1ex]
{\small
$^1$ Max Planck Institute for Solar System Research,
     Justus-von-Liebig-Weg 3, 37077 G\"ottingen, Germany
\\
$^2$ Gesellschaft f\"ur wissenschaftliche Datenverarbeitung,
     Am Fa{\ss}berg 11, 37077 G\"ottingen, Germany
\\
$^3$ Lockheed Martin Solar and Astrophysics Laboratory, 
     Palo Alto, CA 94304, USA
}

\vspace{1ex}

\fancyhead[ER,OR]{}
\fancyhead[EL,OL]{\slshape\footnotesize Nature Physics, ~ final version published online 27 April 2015. ~~ \href{http://dx.doi.org/10.1038/nphys3315}{\color{magenta}DOI: 10.1038/nphys3315}\\
\hspace*{\fill}Main text of this arXiv version is identical to the originally submitted manuscript.}

\begin{multicols}{2}

% ABSTRACT %%%%%%%%%%%%%%%%%%%%%%%%%%%%%%%%%%%%%%%%%%%%%%%%%%%%%%%%%%%%%%%%%%%%

{\bf\noindent
The outer solar atmosphere, the corona, contains plasma at temperatures of more than a million K, more than 100 times hotter that solar surface. How this gas is heated is a fundamental question tightly interwoven with the structure of the  magnetic field in the upper atmosphere. Conducting numerical experiments based on magnetohydrodynamics we account for both the evolving three-dimensional structure of the atmosphere and the complex interaction of magnetic field and plasma. Together this defines the formation and evolution of coronal loops, the basic building block prominently seen in X-rays and extreme ultraviolet (EUV) images.
The structures seen as coronal loops in the EUV can evolve quite differently from the magnetic field. While the magnetic field continuously expands as new magnetic flux emerges through the solar surface, the plasma gets heated on successively emerging fieldlines creating an EUV loop that remains roughly at the same place.
For each snapshot the EUV images outline the magnetic field, but in contrast to the traditional view, the temporal evolution of the magnetic field and the EUV loops can be different. Through this we show that the thermal and the magnetic evolution in the outer atmosphere of a cool star has to be treated together, and cannot be simply separated as done mostly so far.
}

% Intro %%%%%%%%%%%%%%%%%%%%%%%%%%%%%%%%%%%%%%%%%%%%%%%%%%%%%%%%%%%%%%%%%%%%%%%

\begin{figure*}[t]
\noindent
\centerline{\includegraphics[width=0.85\textwidth]{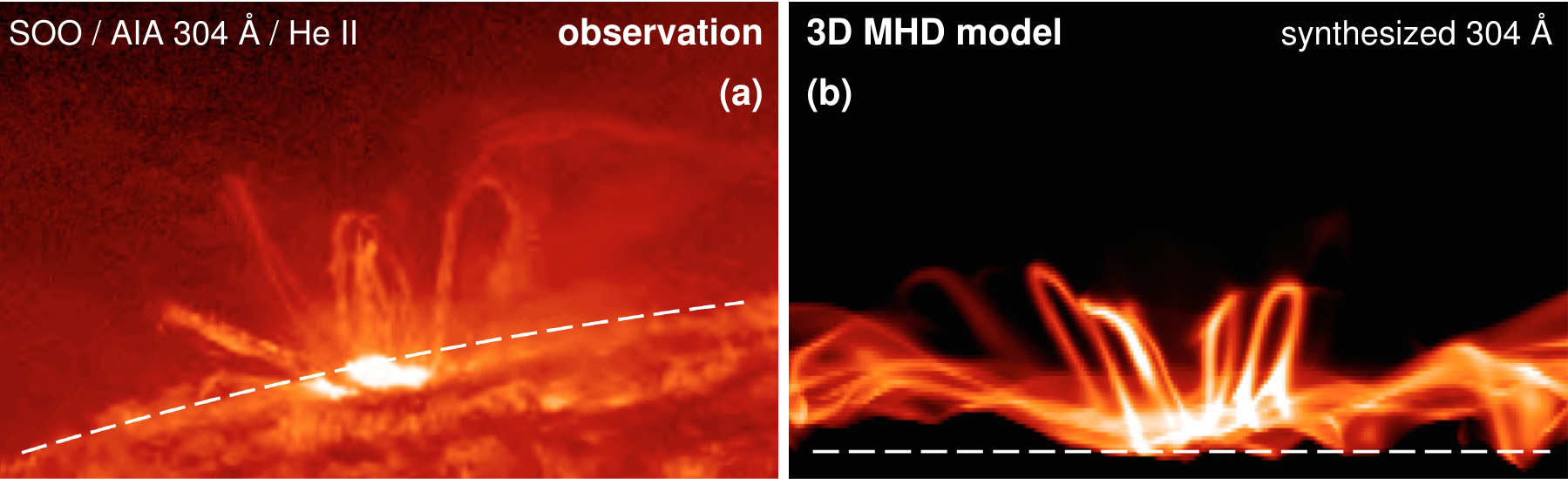}}
\caption{{\bfseries The upper atmosphere of the Sun seen in light emitted by about 100\,000\,K hot plasma}. Panel (a) shows an observation from space with the Solar Dynamics Observatory (SDO) taken in the 304\,\AA\ band dominated by emission from singly ionised He. The limb of the Sun is indicated by the dashed line. Coronal loops are mostly seen edge-on rising some 40\,000\,km above the limb. 
Panel (b) shows a numerical simulation as described in this paper. It shows the synthesised emission in the same 304\,\AA\ channel integrated horizontally through the computational box of a numerical experiment. Similar to the real Sun, loops arch above the surface (dashed line).
\label{F:obs}
}
\end{figure*}

In the upper atmosphere of the Sun the energy density of the magnetic field supersedes the density of the internal or the kinetic energy by far. Thus the magnetic field can easily provide the energy to heat the plasma in the corona to its high temperatures, exceeding those on the surface by more than a factor of 100. The dominance of the magnetic field also gives rise to the coronal loops that are seen so nicely in the extreme ultraviolet (EUV) or X-rays (see \fig{F:obs}): if energy is deposited on a magnetic fieldline, heat conduction in a fully ionised plasma will redistribute that energy efficiently along only that fieldline (but not across). Thus the plasma along a fieldline gets heated and becomes visible in EUV and X-rays. In this picture the EUV and X-ray emission shows the magnetic field in a similar way as iron filings are used in school to show fieldlines of a permanent magnet.
Despite its pivotal importance, actual measurements of the magnetic field in the solar corona are notoriously difficult, and are not yet possible on a regular basis at all places in the upper atmosphere\cite{Peter+al:2012}.

%------------------------------------------------------------------------------
%  FIGURE  1  -----------------------------------------------------------------
%------------------------------------------------------------------------------
%------------------------------------------------------------------------------

% Magnetic field and EUV loops %%%%%%%%%%%%%%%%%%%%%%%%%%%%%%%%%%%%%%%%%%%%%%%%

Instead of direct measurements, mainly extrapolations of the observed magnetic field at the surface provide the magnetic information in the corona\cite{DeRosa+al:2009}. For a comparison, one can combine stereoscopic observations to reconstruct the three-dimensional structure of coronal loops, in particular their path in space\cite{Aschwanden+al:2008}. Comparing such stereoscopy to magnetic field extrapolations provides evidence that the loops seen in EUV indeed outline fieldlines\cite{Feng+al:2007}.  This paradigm underlies both one-dimensional modeling\cite{Mariska:1992,Mikic+al:2013} where the  thermodynamics of the coronal plasma  is treated in detail along assumed static fieldlines, and magnetofrictional modeling\cite{Cheung+DeRosa:2012} where an instantaneous thermal equilibrium is assumed along dynamic fieldlines. On the real Sun we will not find these extreme cases, but a changing magnetic field hosting plasma with an evolving thermal structure. 

% Limitations %%%%%%%%%%%%%%%%%%%%%%%%%%%%%%%%%%%%%%%%%%%%%%%%%%%%%%%%%%%%%%%%%

Models accounting for this three-dimensional structure and evolution of the solar corona already pointed to a mismatch between magnetic and thermal structure\cite{Mok+al:2008}, which plays an important role to understand the cross section of coronal loops\cite{Peter+Bingert:2012}. The thermal evolution, i.e., when plasma gets heated and when a loop becomes visible in EUV, is not coupled to the fieldlines as such, but to the heat input along fieldlines. Thus one can imagine scenarios in which the appearance of coronal loops decouples from the motions of magnetic field lines. We show that such scenarios are realistic for situations on the Sun, and thus our understanding of the structure and evolution of the solar corona, and ultimately the heating processes, will have to fully acknowledge the intimate interaction of the thermal evolution of coronal loops and the changing magnetic structure.

% Model %%%%%%%%%%%%%%%%%%%%%%%%%%%%%%%%%%%%%%%%%%%%%%%%%%%%%%%%%%%%%%%%%%%%%%%

To investigate the relation of the thermal and magnetic evolution in the corona above an active regions we conduct a numerical three-dimensional experiment. For this we solve the problem of magnetohydrodynamics in which the induction equation describing the magnetic field is coupled to the description of a fluid governed by the conservation of mass, momentum and energy. In the latter we account for the heat conduction along the magnetic field, optically thin radiative losses and heating through Ohmic dissipation. Our model follows the philosophy of previous studies in which the magnetic field is driven at the surface of the Sun, which is the lower boundary of the model\cite{Gudiksen+Nordlund:2002,Bingert+Peter:2011,Bourdin+al:2013}. In contrast to these earlier models, we drive our system at the lower boundary by a separate model of an emerging sunspot pair\cite{Rempel+Cheung:2014}. This way coronal loops form in the emerging active region in response to the enhanced Poynting flux into the corona at locations where magnetic field is pushed around, similar to flux braiding\cite{Parker:1972} or flux-tube tectonics\cite{Priest+al:2002}. This new study on the evolution of thermal and magnetic properties is based on the same simulation as used before to investigate the formation of active region loops\cite{Chen+al:2014}.

\fancyhead[EL,OL]{}
\fancyhead[EC,OC]{\slshape\footnotesize Chen, Peter, Bingert \& Cheung: ~ Magnetic Jam in the Corona of the Sun, ~ Nature Physics (2015). ~~ \href{http://dx.doi.org/10.1038/nphys3315}{\color{magenta}DOI: 10.1038/nphys3315}}

% Magnetic field and emission %%%%%%%%%%%%%%%%%%%%%%%%%%%%%%%%%%%%%%%%%%%%%%%%%

In order to study the relation of the magnetic field to the coronal loops seen in emission we have to follow the temporal evolution of both in the simulation. The procedure to follow (a bundle of) magnetic fieldlines as well as an EUV loop is described in the Supplementary Material (SM) \ref{S:follow}. From the output of the MHD simulation, we use the temperature and density at each grid point to evaluate the coronal emission. Integrating along a line-of-sight one then obtains synthetic observations that can be treated as real ones\cite{Peter+al:2006}, which shows the same typical structures as real observations\cite{Bourdin+al:2013}. The example comparison shown in \fig{F:obs} underlines that 3D models now capture the essential observational signatures found for emerging loops, which is a significant step forward in understanding the structure, dynamics and heating of the corona, one of the enigmatic problems in astrophysics. Here we synthesise the coronal emission as it would be seen by the Atmospheric Imaging Assembly (AIA)\cite{Boerner+al:2012} onboard NASA's Solar Dynamics Observatory. For our analysis we synthesise the 193\,\AA\ filter which is dominated by emission from Fe\,{\sc{xii}} that forms at around 1.5 MK, following well established procedures\cite{Peter+Bingert:2012}.

% Evolution of EUV loop ... %%%%%%%%%%%%%%%%%%%%%%%%%%%%%%%%%%%%%%%%%%%%%%%%%%%

\begin{figure*}
\noindent
\centerline{\includegraphics[width=0.85\textwidth]{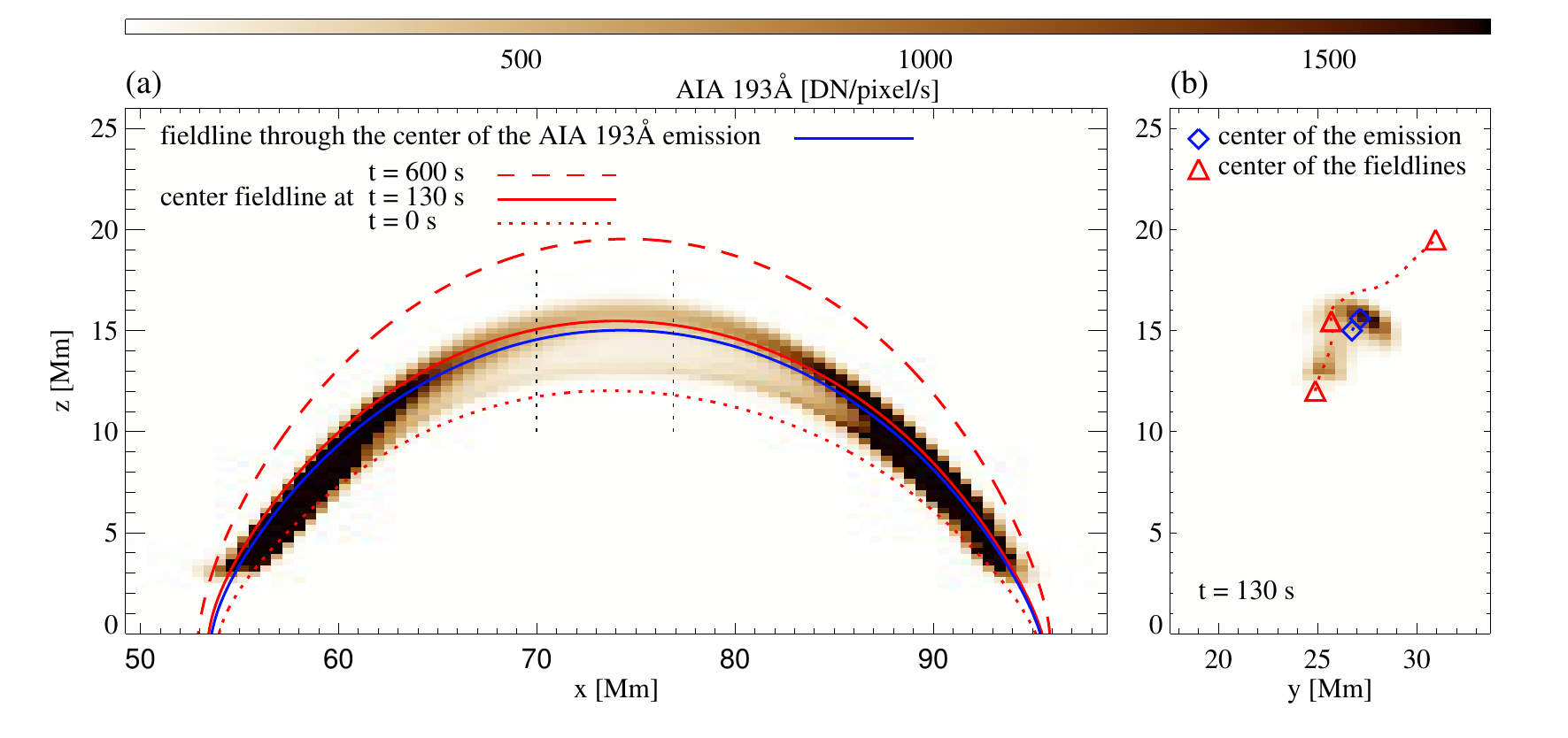}}
\caption{{\bfseries Snapshot of the coronal loop in the numerical simulation.} This shows synthesised emission as seen in a wavelength band at 193\,\AA\ dominated by Fe\,{\sc{xii}} forming near 1.5\,MK.
Panel (a) displays the loop from the side with the emission integrated through the computational box at time 130\,s. The emission pattern remains more or less at the same place (cf.\ \fig{F:time.series} in SM~\ref{S:loop}). In contrast, the fieldlines expand, here indicated by the \emph{same} fieldline shown at three different times (0\,s, 130\,s, and 600\,s). For comparison the blue line shows the fieldline through the center of the emission structure at 130\,s (see SM~\ref{S:follow} for a more precise definition of the red and blue lines).
To get a better impression of the 3D structure, panel (b) shows the middle part of the loop integrated along the loop (from $x{=}$70\,Mm to 77\,Mm as indicated by the dotted lines in panel a). Here the image shows again the 193\,\AA\ channel emission, the blue diamond the center of the EUV loop and the red triangles the position of the fieldline in the $x$=74\,Mm plane at the same three times as in panel (a).
These plots cover only part of the computational domain (${\approx}~$150 ${\times}$ 75 ${\times}$ 50~Mm$^3$ in the $x$, $y$,and $z$ directions).
%\newline
A movie showing the temporal evolution is available in the online edition.
\newline
It is also available at 
%\newline
\href{http://www2.mps.mpg.de/data/outgoing/peter/papers/2015-magnetic-jam/movie-fig-2.mp4}{\color{magenta}http://www2.mps.mpg.de/data/outgoing/peter/papers/2015-magnetic-jam/movie-fig-2.mp4}
\label{F:side}
}
\end{figure*}

In \fig{F:side} we show the synthesised 193\,\AA\ observation when integrating horizontally through the computational domain. The snapshot shown here clearly shows a coronal loop hosting million K hot plasma. Following the temporal evolution in the movie that is available in the online edition (further snapshots in \fig{F:time.series} in SM \ref{S:loop}) it is evident that the EUV loop forms, becomes bright and then starts fading over the course of a good fraction of an hour. Most importantly, the EUV loop, i.e., the pattern visible in the 193\,\AA\ channel, remains at more or less the same place. In particular the EUV loop is \emph{not} expanding upwards.

%------------------------------------------------------------------------------
%  FIGURE  2  -----------------------------------------------------------------
%------------------------------------------------------------------------------
%------------------------------------------------------------------------------

% ... and fieldlines %%%%%%%%%%%%%%%%%%%%%%%%%%%%%%%%%%%%%%%%%%%%%%%%%%%%%%%%%%

This is in contrast to the evolution of the structure of the magnetic field. Also in \fig{F:side}, we overplot \emph{one single} magnetic fieldline at different times. It can be seen that this fieldline moves upwards while the active region is emerging (see the movie or \fig{F:time.series} in SM \ref{S:loop}). In \fig{F:side} (and the movie) we also show the coronal emission in a vertical slab in the middle of the loop (and perpendicular to the loop) to emphasise how differently the pattern of the EUV emission evolves compared to the magnetic structure. There is no mass flow across fieldlines. We emphasise that at each snapshot the EUV loop is roughly following a fieldline, but at each time it is a different fieldline that is aligned with the EUV loop.

% fieldlines, heating and emission %%%%%%%%%%%%%%%%%%%%%%%%%%%%%%%%%%%%%%%%%%%%

To understand this behaviour one has to investigate the heat input and the resulting temperature, density and emission structure along individual fieldlines, the details of which are described in SM \ref{S:loop}. We find that each individual fieldline shows an increased Ohmic heating for about one hundred seconds. This heats up the plasma and through evaporation of gas from the lower atmosphere the density of the loop increases. Because the EUV passbands are sensitive to a limited range of temperatures only, the plasma along each expanding fieldline is brightening up only for some 50\,s to 100\,s. The fieldlines get heated in succession, i.e., one after the other while moving upwards. This creates a more or less stationary pattern of increased emission, while the structure of the magnetic field is constantly moving upwards. This might be compared to a traffic jam triggered by at a construction site on a highway. Here all cars (defining the structure) are moving forward, but the construction site (heating up not the cars but the tempers) and thus the pile-up of cars (defining the pattern) remain at the same location.

% tectonics in the photosphere %%%%%%%%%%%%%%%%%%%%%%%%%%%%%%%%%%%%%%%%%%%%%%%%

\begin{figure*}[t]
\noindent
\centerline{\includegraphics[width=0.85\textwidth]{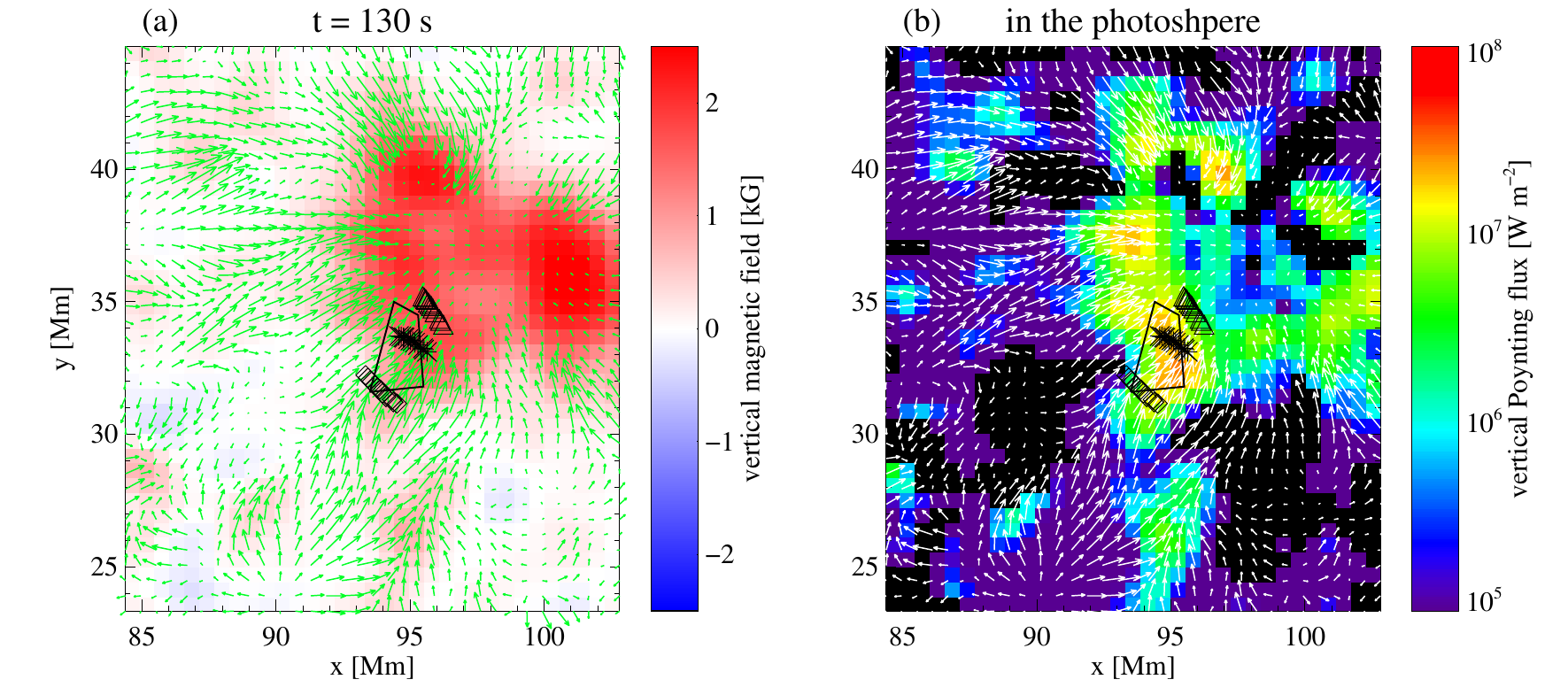}}
\caption{{\bfseries Evolution at the solar surface while the coronal loop forms.} 
Panel (a) shows the vertical magnetic magnetic field and panel (b) the vertical component of the Poynting flux, both in the photosphere. These snapshots are taken at the time $t{=}$130\,s. 
The concentration of magnetic field in panel (a), seen in red, shows the location of one of the two sunspots that form the active region in this simulation.
For the time $t{=}-$600\,s we indicate a number of positions by the diamonds that are located at the footpoints of fieldlines that transverse the bright coronal loop later. The asterisks and triangles show the position of these locations at later times $t{=}+$130\,s, and $+$1200\,s, when they are carried with the coalescent flow forming the sunspot.
The field-of-view covers only a small fraction of the whole computational domain (${\approx}~$150 ${\times}$ 75~Mm$^2$ in the horizontal directions). 
\newline
An animation showing the temporal evolution over 30 minutes from $t{=}-$600\,s to $+$1200\,s is available in the online edition.
\newline
It is also available at
%\newline
\href{http://www2.mps.mpg.de/data/outgoing/peter/papers/2015-magnetic-jam/movie-fig-3.mp4}{\color{magenta}http://www2.mps.mpg.de/data/outgoing/peter/papers/2015-magnetic-jam/movie-fig-3.mp4}
\label{F:photosphere}
}
\end{figure*}

The reason for the transient enhancement of the heating along individual fieldlines is found at their roots in the photosphere. The coalescent flow that forms the sunspot drives magnetic patches towards the strong magnetic field of the sunspot\cite{Rempel+Cheung:2014,Cheung+al:2010}. This is shown by the arrows in \fig{F:photosphere} that display the horizontal flow field in the photosphere.
Consequently, at the outer edge of the spot,  there will be a region of enhanced (vertical) Poynting flux, i.e., the upward directed flux of magnetic energy. This is similar to the flux-tube-tectonics model\cite{Priest+al:2002} where (horizontal) shuffling of magnetic patches leads to an upward directed flux of magnetic energy, which is then available to heat the coronal plasma.
Because each fieldline is pushed into the spot and thus transverses the region of the enhanced Poynting flux, the heat input into the corona along individual fieldlines is transient (see detailed discussion in SM \ref{S:loop} and movie attached to \fig{F:poynting}). In short, one would expect the coronal EUV loops to show up wherever there are strong (horizontal) gradients of the magnetic field at the footpoints, which are the locations of strong currents and hence strong Poynting flux.

%------------------------------------------------------------------------------
%  FIGURE  3  -----------------------------------------------------------------
%------------------------------------------------------------------------------
%------------------------------------------------------------------------------

% cartoon summary %%%%%%%%%%%%%%%%%%%%%%%%%%%%%%%%%%%%%%%%%%%%%%%%%%%%%%%%%%%%%%

\begin{figure*}[b]
\noindent
\begin{minipage}[b]{\columnwidth}
\centerline{\includegraphics[width=0.9\textwidth]{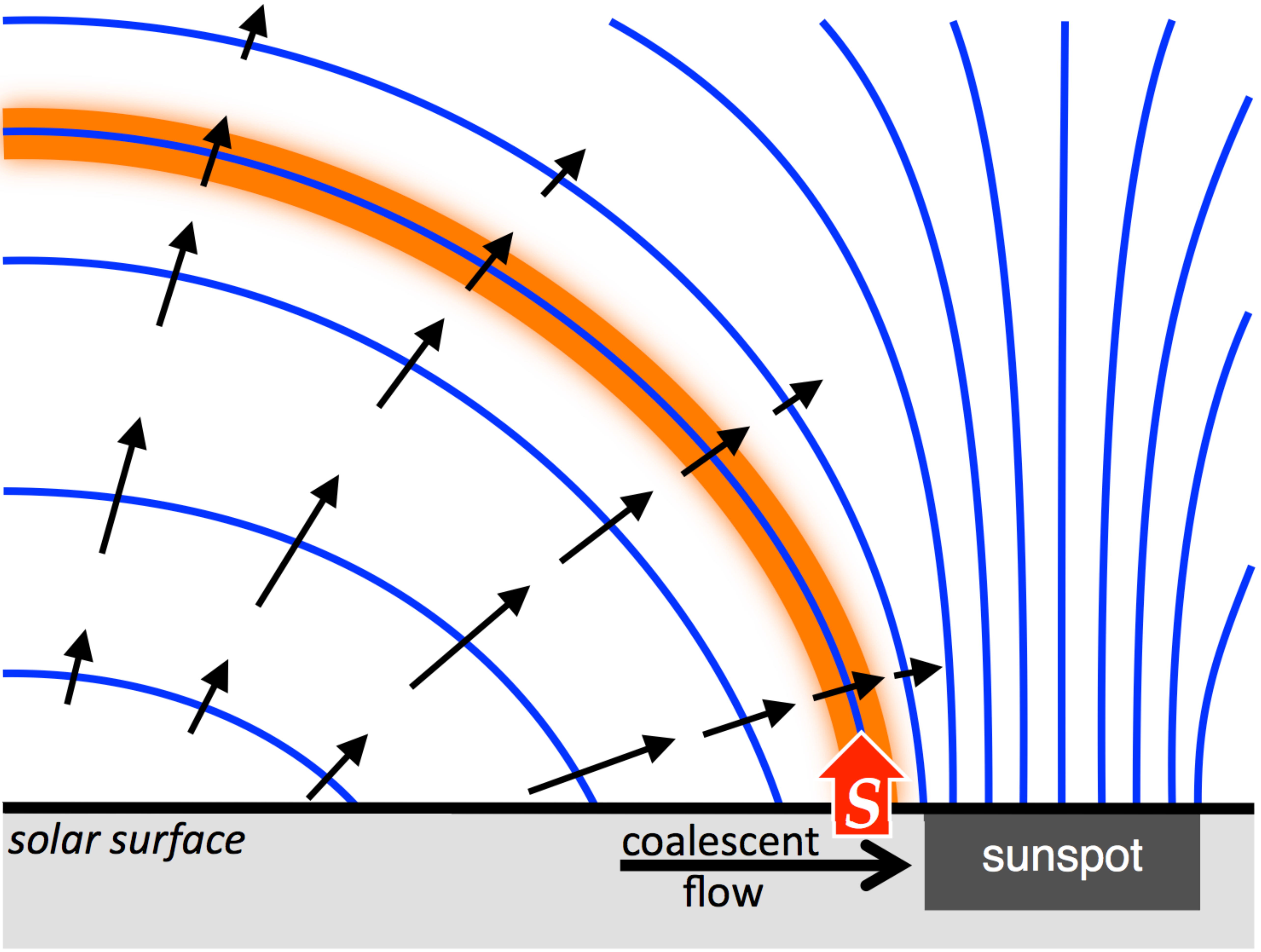}}
\end{minipage}
\hfill
\begin{minipage}[b]{\columnwidth}
\caption{{\bfseries Cartoon showing the interplay between magnetic field expansion and the EUV loop.} A coalescent flow forming the sunspot drags the magnetic field in the photosphere near the solar surface into the sunspot. In response a hot spot of enhanced upward directed Poynting flux, $S$, forms (red arrow). The expanding fieldlines (blue) move upwards and to the side. When they transverse the hot spot of Poynting flux, the plasma on that fieldline gets heated and brightens up. As the fieldline expands further, it leaves the hot spot and gets darker again. In consequence a bright coronal EUV loop forms (orange) and remains rather stable while the successively heated fieldlines move through.
\label{F:cartoon}
}
\end{minipage}
\end{figure*}

This mechanism is illustrated by the cartoon in \fig{F:cartoon}. During the emergence of magnetic flux forming a sunspot pair the field is pushed upwards and to the sides. In sunspots, where the magnetic field is very strong, convection is suppressed, and thus the flows driving the coalescence of the magnetic field come to a halt. Whenever a fieldline is crossing the region of the enhanced Poynting flux, energy is deposited along that fieldline and the plasma on it is heated. Consequently this fieldline becomes visible in EUV for a short time. With successive fieldlines passing the "hot spot" of Poynting flux they all brighten roughly at the same place, creating the illusion of a static emission pattern forming a loop, while the magnetic field is moving.

%------------------------------------------------------------------------------
%  FIGURE  4  -----------------------------------------------------------------
%------------------------------------------------------------------------------
%------------------------------------------------------------------------------

In our 3D numerical experiment we find that the temporal evolution of the structure of the magnetic field in the corona of the Sun can be radically different from that of the patterns seen in the coronal emission.  This implies that modeling the \emph{temporal evolution} of EUV loops as 1D structures following a fieldline is a problematic concept in regions where the magnetic field is evolving, i.e. whenever the Sun gets dynamic --- and interesting. Thus the credibility of many of the time-dependent 1D loop models that have been used as the workhorse in theoretical coronal  studies over the last two decades need to be reconsidered. Still,  at
any given snapshot the coronal EUV loops in our model outline magnetic field lines. Therefore EUV observations should provide useful when implemented into procedures to recover a snapshot of the coronal magnetic field through extrapolation methods\cite{Malanushenko+al:2014}. In summary, we have to treat the magnetic and the thermal evolution of the corona as one single problem --- this requires to have a more holistic view of the magnetic and thermal properties of the corona when addressing the question of the structure, dynamics and heating of the corona.

%%%%%%%%%%%%%%%%%%%%%%%%%%%%%%%%%%%%%%%%%%%%%%%%%%%%%%%%%%%%%%%%%%%%%%%%%%%%%%%
%\clearpage

\vspace{8ex}

\noindent
{\bfseries References}

\vspace{2ex}

{\small\noindent

%\bibliography{all,local}
%\bibliographystyle{naturemag}

}

{\small\noindent\footnotesize
References 21--38 are cited in the Supplementary Material only.
}

%\clearpage

\vspace{5ex}

{\small\noindent

{\bf\noindent  Correspondence}
\\ 
Correspondence and requests for materials
should be addressed to H.P.~(email: peter@mps.mpg.de).

\vspace{3ex}

{\bf\noindent  Acknowledgements}
\\
We thank R.~Cameron for discussions and comments on the manuscript. This work was supported by the International Max-Planck Research School (IMPRS) for Solar System Science at the University of Göttingen. It was was partially funded by the Max-Planck/Princeton Center for Plasma Physics. The computations were done at GWDG in G\"ottingen and on SuperMUC in at LRZ in Munich. We acknowledge PRACE for awarding us the access to SuperMUC based in Germany at the Leibniz Supercomputing Centre (LRZ).

\vspace{3ex}

{\bf\noindent  Individual contributions to the paper} 
\\
The numerical experiment was designed by H.P. and S.B, the numerical simulation has been conducted by F.C. and S.B., the analysis of the data has been done by F.C., H.P., and S.B., the boundary conditions have been provided and implemented by M.C. and F.C., and H.P. and F.C. wrote the text.

}

%%%%%%%%%%%%%%%%%%%%%%%%%%%%%%%%%%%%%%%%%%%%%%%%%%%%%%%%%%%%%%%%%%%%%%%%%%%%%%%

\end{multicols}

\clearpage
%===============================================================================
%===============================================================================
%===                        ====================================================
%=== SUPPLEMENTARY MATERIAL ====================================================
%===                        ====================================================
%===============================================================================
%===============================================================================

\newgeometry{text={181.5mm,243mm},centering}
\setlength{\columnsep}{6mm}

\setcounter{page}{1}
\setcounter{figure}{0}

\renewcommand{\thefigure}{S\arabic{figure}}
\renewcommand{\thesection}{S\arabic{section}}
\renewcommand{\thepage}{SM~~\arabic{page}}

\newcommand{\figureskip}{\vspace{2ex}\par
                         \centerline{\rule{0.8\textwidth}{0.5pt}}
                         \par\vspace{1ex}}

\newcommand{\mysubsection}[1]{\vspace{2ex}\noindent{\bfseries{#1}}\par\vspace{1ex}\noindent}

\newcommand{\formatting}[1]{#1}

\captionsetup[figure]{labelfont={sf,bf,small},textfont={rm,small}}

\fancyhead[EC,OC]{}
\fancyhead[EL,OL]{\slshape\footnotesize Nature Physics, ~ final version published online 27 April 2015. ~~ \href{http://dx.doi.org/10.1038/nphys3315}{\color{magenta}DOI: 10.1038/nphys3315}}

\noindent
%------------------------------------------------------------------------------
{\large\sf\bfseries Supplementary Material for} 
%------------------------------------------------------------------------------
\\[3ex]
{\Large\sf\bfseries 
Magnetic Jam in the Corona of the Sun
}
\\[2ex]
{
F.~Chen,$^{1}$ 
H.~Peter,$^{1}$ 
S.~Bingert,$^{2}$ 
M.C.M.~Cheung$^{3}$
}
\\[1ex]
{\small
$^1$ Max Planck Institute for Solar System Research,
     Justus-von-Liebig-Weg 3, 37077 G\"ottingen, Germany
\\
$^2$ Gesellschaft f\"ur wissenschaftliche Datenverarbeitung,
     Am Fa{\ss}berg 11, 37077 G\"ottingen, Germany
\\
$^3$ Lockheed Martin Solar and Astrophysics Laboratory, 
     Palo Alto, CA 94304, USA
}

\vspace{2ex}

\noindent
{\bfseries
\ref{S:model}~~ Three-dimensional coronal models compared to observational properties.
\\[1ex]
\ref{S:follow}~~ Following (a bundle of) magnetic field lines and an EUV loop in time.
\\[1ex]
\ref{S:loop}~~ Thermal evolution and coronal emission along individual fieldlines.
}

\vspace{3ex}

{\small
\noindent
Movies for the evolution of the snapshots displayed in \figs{F:side} and \ref{F:photosphere} of the main text and in \fig{F:poynting} of the supplementary material are available in the online edition or at
\href{http://www2.mps.mpg.de/data/outgoing/peter/papers/2015-magnetic-jam/}{\color{magenta}http://www2.mps.mpg.de/data/outgoing/peter/papers/2015-magnetic-jam/}.
}

\vspace{3ex}

%------------------------------------------------------------------------------
\section{\large Three-dimensional coronal models compared to observational properties\label{S:model}}
%------------------------------------------------------------------------------

\begin{multicols}{2}

\mysubsection{Heat input in the model}%
Our three-dimensional (3D) model of the corona above an emerging active region is driven by the convective motions in the photosphere which are prescribed at the lower boundary. To set the velocity field as well as the density, temperature, and magnetic field at the bottom boundary of our computational box, we use the time-dependent output of a model of a magnetic fluxtube emerging through the surface\cite{Rempel+Cheung:2014,Cheung+al:2010}. The upper boundary of that flux-emergence model is located about 1\,Mm above the solar surface (where the optical depth is unity) and thus does not include the evolution of the corona above. We use the conditions of the flux-emergence model at the solar surface to prescribe the lower boundary condition of our coronal model.

The horizontal motions at the bottom boundary together with the magnetic field produce a Poynting flux that on average is upward directed. This is similar to the concept of fieldline braiding\cite{Parker:1972} and fluxtube tectonics\cite{Priest+al:2002} (see also SM~\ref{S:loop} for the Poynting flux at the coronal base). The disturbances of the magnetic field propagate upwards, the currents induced by this are dissipated, and consequently magnetic energy is converted to internal energy through Ohmic heating.

When solving the basic equations of magnetohydrodynamics (MHD)\cite{Priest:2014}, we assume a constant value of the magnetic resistivity $\eta$ in the induction equation and the energy equation. For $\eta$ we choose a value so that the magnetic Reynolds number $R_m=UL/\eta$ is of order unity when we choose the grid spacing as the length scale $L$ (and the sound speed as the typical velocity $U$). This ensures that the magnetic energy is mostly dissipated at the smallest scales resolved by the numerical simulation and that at the same time the numerical dissipation is small compared to the Ohmic dissipation. 
In the present simulation we used $\eta=10^{10}$\,m$^2$/s.
Previous studies have shown that for such a setup the energy input into the system will be independent of the choice of $\eta$, as long as $\eta$ is adapted to the grid spacing of the simulation\cite{Hendrix+al:1996,Galsgaard+Nordlund:1996,Rappazzo+al:2008}.
To perform the numerical simulation we employed the Pencil code\cite{Brandenburg+Dobler:2002,Bingert+Peter:2011} (http://pencil-code.googlecode.com/).

One particular feature of the energy input of this model is that there is a continuous distribution of heat deposition into the upper solar atmosphere, all the way from the chromosphere to the corona. Typically the energy input per volume through Ohmic heating shows a monotonic decrease with height. In the coronal part this decrease is roughly exponential with a scale height ranging from some 5\,Mm to 10\,Mm, depending on the structures on the surface. This applies to the average drop\cite{Gudiksen+Nordlund:2002} as well as for the variation along individual fieldlines\cite{vanWettum+al:2013}. The heating per particle peaks in the transition region from the chromosphere to the corona\cite{Hansteen+al:2010,Bingert+Peter:2011} because the scale height of the heating is in-between the pressure scale heights of the chromosphere and the corona.

\formatting{\vspace{0.7ex}}

\mysubsection{Comparison to observational properties}%
The type of model used here has been employed for a decade and successfully reproduced and explained a number of puzzling observations in the solar corona. Using spectral line profiles synthesised from 3D models of the corona above an active region, the comparison of the \emph{average} properties of model and observations showed a good match. This applies to the redshifts in the transition region and of the average differential emission measure\cite{Peter+al:2004,Peter+al:2006}, as well as to the temporal fluctuations of Doppler shift and intensity\cite{Peter:2007}. A later 3D model gave an explanation also of the coronal blueshifts\cite{Hansteen+al:2010}. Investigations of individual loops appearing in the emission synthesised from the model show an intensity variation along the loop comparable to observations, and a lifetime comparable to observed solar loops\cite{Peter+Bingert:2012}. These studies also provided a new explanation for the puzzling observational finding that many loops seem not to expand with height\cite{Peter+Bingert:2012,Chen+al:2014}. Furthermore, the 3D models show envelopes of groups of loops similar to observations\cite{Peter+al:2013.hic}. Many of the loops appearing in the models have densities higher than expected from hydrostatic equilibrium, just as found in observations\cite{Bourdin+al:2014}

The above models have not been fine-tuned to reproduce specific features on the Sun; sample loops from the models have been picked to compare them to actual observation, or average properties between the 3D model and observations have been compared. 
In one particular numerical experiment a setup was chosen in which the magnetic field and the horizontal velocities at the bottom boundary of the model were taken from an actual observation of an active region in the photosphere. The comparison of the resulting coronal structures shows that the loops forming in the model appear at the same locations as the coronal loops have been observed on the real Sun, which was confirmed by a 3D analysis of stereoscopic solar observations\cite{Bourdin+al:2013}.

\end{multicols}

\fancyhead[EL,OL]{}
\fancyhead[EC,OC]{\slshape\footnotesize Supplementary Material for ~ Chen, Peter, Bingert \& Cheung,  ~ Nature Physics (2015). ~~ \href{http://dx.doi.org/10.1038/nphys3315}{\color{magenta}DOI: 10.1038/nphys3315}}

\vspace{8ex}

%------------------------------------------------------------------------------
\section{\large Following (a bundle of) magnetic field lines and an EUV loop in time\label{S:follow}} 
%------------------------------------------------------------------------------

\begin{multicols}{2}

Conventionally, it is agreed that coronal loops seen in the EUV and X-ray images approximately indicate the magnetic field lines. This is because the magnetic energy is dominating in the corona, the emitting plasma is confined by magnetic field, and the heat conduction is very sufficient along magnetic field lines. Stereoscopic observations\cite{Feng+al:2007} and 3D simulations\cite{Gudiksen+Nordlund:2005a, Gudiksen+Nordlund:2005b, Bingert+Peter:2011} of active regions with a gradual evolution of the (photospheric) magnetic field support this concept. However, if the magnetic field would be dynamic, for example, during the formation of an active region, this concept has to be questioned, in particular if the timescales of the magnetic evolution and the thermal evolution of the plasma would be different.

 To investigate the spatial relation between coronal loops and magnetic field lines, we need to follow the field lines in time. For this we assume that plasma elements are frozen-in to the magnetic field lines, which is true if the magnetic Reynolds number is large, and which is the case in most of the corona. Then we follow the motion of a selected plasma element and calculate the magnetic field line through the plasma element at each instant time. In our numerical model, we have to employ a certain magnetic diffusivity, which unfortunately allows plasma to move across the magnetic field lines. However, the average diffusion speed across 10\,Mm is of the order of 1\,km/s. This is much smaller than the typical perpendicular velocity (of well above 10\,km/s) due to the expansion of the magnetic field, as will be detailed below.

The snapshot cadence of the numerical simulation is 30\,s, which sufficiently captures the evolution of MHD variables in our simulation. When we follow the field lines as outlined above, the selected plasma element might move across several grid points in one time step. However, because the evolution of the MHD variables is smooth, we can safely use a cubic spline method to interpolate the snapshots to a sufficiently high cadence of 1\,s and use this to follow the gas packages on the fieldlines. We tested this interpolation for part of the time series by writing snapshots with 1\,s cadence and found the same results.

%\formatting{\vspace{2ex}}

\mysubsection{Axis of a magnetic tube}%
To follow the magnetic field, we select twelve points in the vertical middle plane of the simulation box as seeds. This plane is in the middle between the two sunspots in the photosphere and perpendicular to the connecting line between the spots. Thus it is also roughly perpendicular to the EUV loops that form and connect the opposite polarities of the emerging active region. Of these seeds, eleven form a circle of roughly 2\,Mm in diameter and the twelfth is in the middle of that circle. The initial selection of these twelve points is chosen so that at some time ($t{=}$130\,s) these points roughly encircle the EUV loop that forms. Basically the 11 points define a magnetic tube and the twelfth point is on the axis of that tube.

We follow these fieldlines in time (backwards and forwards).
First we trace each fieldline from each of the initial points. Then we follow the fieldline in time by assuming that the fieldlines are frozen-in with the gas. In practice, we follow the gas parcel near the apex of the fieldline using the plasma velocity at that location, and then use the new location of that gas parcel at the next time step as the new point for tracing the fieldline at that next time step. This is done forward and backward in time until the time period of interest is covered. The electric conductivity in the model is sufficiently high so that the (numerical) diffusion speed of the plasma through the magnetic field is small compared to the actual speed of the rising magnetic field lines.
For each of the fieldlines we calculate the positions of their respective intersections with the middle plain, $\vec{r}_{i}$, where $i$ is the index from 1 to 12. The position of the center of the magnetic tube in the middle plan, $\vec{c}_{\rm{mag}}$, we define as 
$$
\vec{c}_{\rm{mag}} = \frac{1}{12}\sum_{i=1}^{12} \vec{r}_{i} ~.
$$
Tracing the field line for each time step from this point provides us with the fieldline of the axis of the magnetic tube. We chose this procedure because the magnetic tube will change its shape while expanding. In general, a tube with a circular cross section will get deformed into a more elongated (or even more strangely shaped) cross section\cite{Chen+al:2014}. By using the axis of the magnetic tube we get a better representation of the evolution of the magnetic tube independent of the shape of the cross section of the tube. The center fieldlines plotted in red color in \fig{F:side} in the main text and its attached movie, as well as in \fig{F:time.series} are these axis of the magnetic tube.

The vertical speed associated with the upward expansion of the apex of the fieldline is about 30\,km/s, as can be seen by inspection of the movie attached to \fig{F:side} in the main text (or in \fig{F:time.series}).

\mysubsection{Axis of an EUV loop}%
To follow (the axis of) the EUV loop we use the emission synthesised in the 193\,\AA\ band as it would be observed with AIA\cite{Lemen+al:2012}. This shows plasma at temperatures of about 1.5\,MK. We calculate the center-of-gravity of the emission in the vertical midplane and calculate the magnetic fieldline through this point. 

If the emission at each gridpoint in the midplane is $\varepsilon_i$, and the position of that gridpoint is $\vec{r}_{i}$, then the center-of-gravity of the emission is
$$
\vec{c}_{\rm{emiss}} =\frac{\sum_i \varepsilon_i\vec{r}_{i}}{\sum_i\varepsilon_i} ~.
$$
For convenience (with no impact on the result) we carry out the summations only over those gridpoints with an emissivity above a certain threshold (20\,DN/pixel/s/Mm).

For each timestep we now calculate this center position of the EUV loop in the midplane and follow the magnetic fieldline through it. This fieldline we define as the axis of the EUV loop (plotted in blue color in \fig{F:side} in the main text and its attached movie, as well as in \fig{F:time.series}).

\end{multicols}

%%%%%%%%%%%%%%%%%%%%%%%%%%%%%%%%%%%%%%%%%%%%%%%%%%%%%%%%%%
%%%%%%%%%%%%%%%%%%%%%%%%%%%%%%%%%%%%%%%%%%%%%%%%%%%%%%%%%%
%%%%%%%%%%%%%%%%%%%%%%%%%%%%%%%%%%%%%%%%%%%%%%%%%%%%%%%%%%

%\clearpage

\vspace{7ex}

%------------------------------------------------------------------------------
\section{\large Thermal evolution and coronal emission along individual fieldlines\label{S:loop}} 
%------------------------------------------------------------------------------

\begin{multicols}{2}

In order to understand the difference in temporal evolution of the EUV structures and the magnetic field lines we first investigate the actual heat input on individual fieldlines. For this we use the fieldlines mentioned in SM~\ref{S:follow}. We then study how this heat input establishes the density and temperature structure along the fieldlines.  Eventually, this sets the EUV emission along the fieldlines. Finally by relating the evolution of the fieldlines to the Poynting flux at the coronal base we can understand what causes the magnetic field to apparently move through the EUV loop.

For the study of the temporal evolution we choose an arbitrary  zero time, $t{=}$0. At this time the loop as seen in EUV is just about to form. We use this zero time throughout the manuscript, so negative times refer to the temporal evolution before the EUV loop formed. All the times given in figures and movies are with respect to this zero time.

%------------------------------------------------------------------------------
\begin{figure*}[t]
\noindent
\begin{minipage}[b]{\columnwidth}
\includegraphics[width=\textwidth]{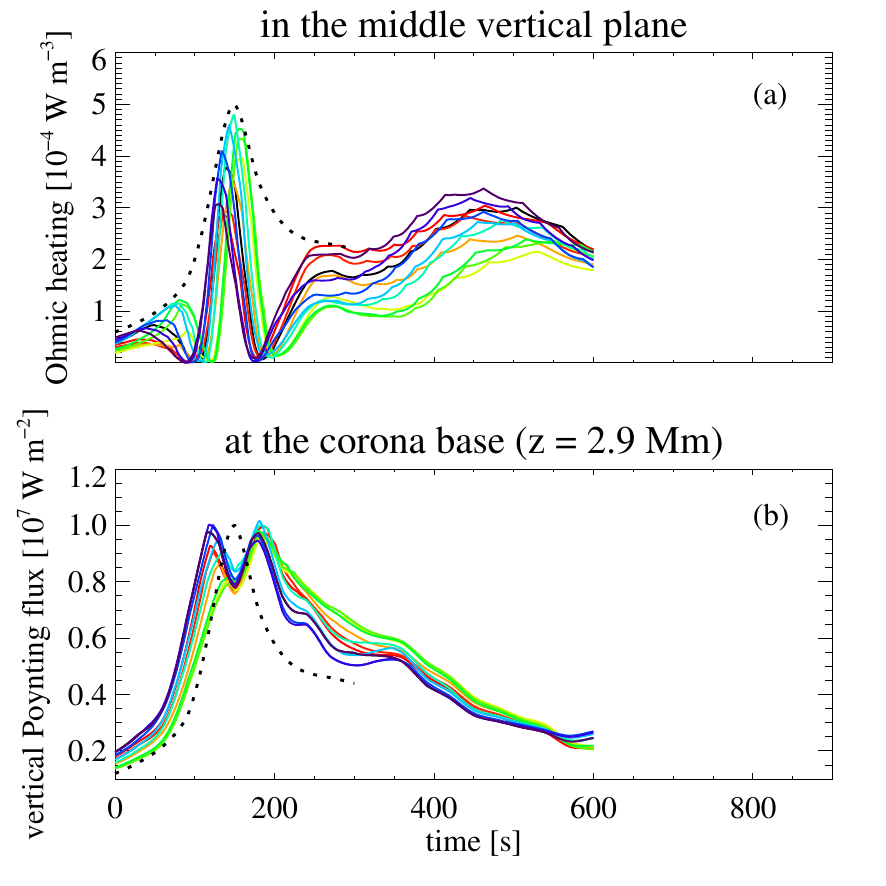}
\end{minipage}
\hfill
\begin{minipage}[b]{\columnwidth}
\caption{
Heating along individual field lines. The coloured lines show the temporal variation of the heating for the twelve fieldlines as defined in SM~\ref{S:follow}.
Panel (a) shows the volumetric energy deposition due to Ohmic dissipation at the cross section of the respective field line and the vertical midplane between the footpoints. This is close to the apex of the respective fieldline. The dotted line displays an envelope for the heat input.
Panel (b) shows the magnetic energy flux into the loop, viz. the vertical component of the Poynting flux as defined in \equ{E:Poynting.vertical}, at the base of the corona. For comparison the dotted envelope from panel (a) is plotted in panel (b), too, just scaled to roughly match the peak of the Poynting flux.
\label{F:heating}
}
\end{minipage}
\vspace{6.5ex}
\end{figure*}
%------------------------------------------------------------------------------

\mysubsection{Heat input for individual fieldlines}%
The density and temperature structure along each fieldline is set by the heat input. To describe the temporal evolution of the heat input we investigate two aspects, the volumetric heat input due to Ohmic dissipation near the loop apex, and the flux of magnetic energy into the loop at the coronal base.

For the volumetric heating we investigate the Ohmic heating of the crossing point of the respective fieldline with the vertical midplane between the two sunspots. Because the setup is quite symmetric it is close (but not identical) to the heat input at the apex of the fieldline. Following the fieldline in time as outlined in SM~\ref{S:follow} we find the time variation of the heat input near the apex for that particular fieldline. In \fig{F:heating}a we show this for the twelve fieldlines discussed in SM~\ref{S:follow} that coincide with the bright EUV loop at $t{\approx}$130\,s. It is clear that each fieldline is heated for some 50\,s, with all the fieldlines defining the magnetic tube of roughly 2\,Mm diameter peaking over times from $t{\approx}$130\,s to 160\,s.

\begin{figure*}[t]
\noindent
\begin{minipage}[b]{\columnwidth}
\includegraphics[width=\textwidth]{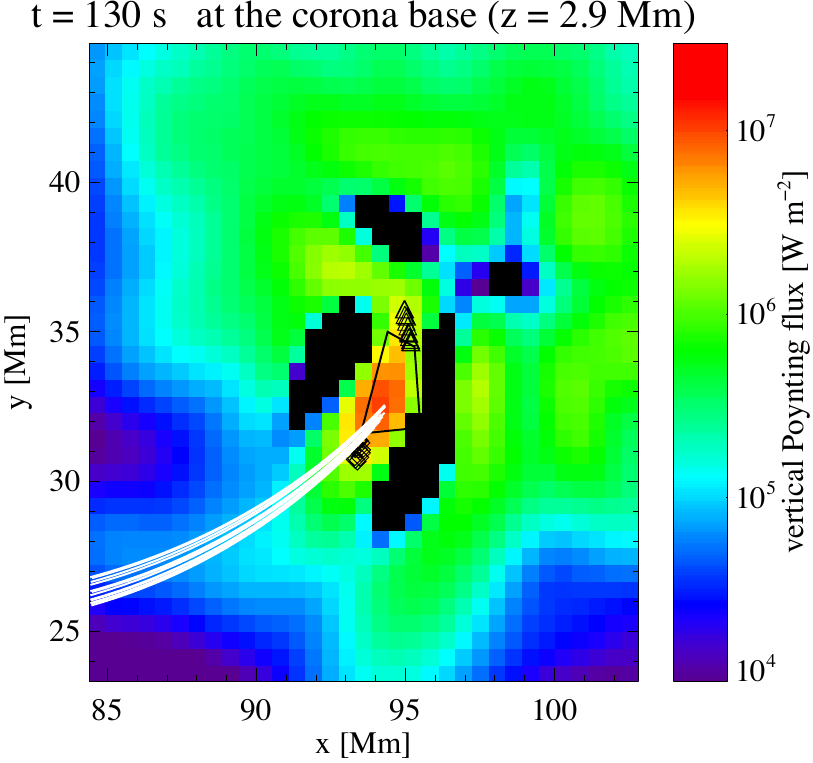}
\end{minipage}
\hfill
\begin{minipage}[b]{\columnwidth}
\caption{
Hot spot of Poynting flux at the base of the corona. The image shows the vertical Poynting flux at the base of the corona as defined in \equ{E:Poynting.vertical}. The red region in the middle of the image shows the location of the enhanced Poynting flux. This is the hot spot of energy flux into the corona. The white lines show the projection of the magnetic fieldlines of the magnetic tube defined in SM~\ref{S:follow} at time $t{=}$130\,s. The diamonds and the triangles indicate the position of the footpoints of these fieldlines at times $t{=}$0\,s and 600\,s. The pattern of the Poynting flux remains rather stable over 10\,minutes. The field-view and the polygon roughly encircling the hot spot are the same as in \fig{F:photosphere} in the main text.
\newline
The temporal evolution is shown in a movie available online.
\newline
The movie is also available at:
\newline
%\href[options]{URL}{text}
\href{http://www2.mps.mpg.de/data/outgoing/peter/papers/2015-magnetic-jam/movie-fig-s2.mp4}{\color{magenta}http://www2.mps.mpg.de/data/outgoing/peter/papers/2015-magnetic-jam/movie-fig-s2.mp4}
\label{F:poynting}
}
\end{minipage}
\vspace{6.5ex}
\end{figure*}

To investigate the flux of magnetic energy into the coronal part of the fieldline we study the vertical Poynting flux at be base of the corona for each fieldline. For simplicity we use the height of $z{=}$2.9\,Mm, which is the average height where the temperature rises above 10$^5$\,K. In general the Poynting flux is defined as
$ %\begin{equation}\label{E:Poynting.general}
S= \eta\,j{\times}B - (v{\times}B){\times}B/\mu_0 ,
$ %\end{equation}
with the current $j$, the magnetic field $B$, velocity $v$, magnetic resistivity $\eta$, and the magnetic permeability $\mu_0$. At the base of the corona, where the magnetic energy density already dominates the thermal energy density, the first term involving the currents is negligible. The $(v{\times}B){\times}B$ term contains the contribution from emerging horizontal fields and (almost) vertical fields being shifted around, e.g., following the concept of braiding\cite{Parker:1972} or the tectonics\cite{Priest+al:2002}. Because we consider only the energy input into the fieldlines reaching coronal heights, we consider only the latter part.
Finally, we are left with the contribution to the vertical Poynting flux involving the velocity $v_\perp$ perpendicular to the magnetic field. So in terms of the components along the horizontal $x$- and $y$-directions and the vertical $z$-direction the vertical Poynting flux is
\\[0.5ex]
\begin{equation}\label{E:Poynting.vertical}
S_z = ~-~\frac{1}{\mu_0}\Big(v_{{\perp}x}\,B_x~+~v_{{\perp}y}\,B_y\Big)B_z .
\end{equation}
\\[0.5ex]
We evaluate this quantity for each of the fieldlines at the base of the corona (at each of its legs). 
The temporal evolution of this quantity (for the ``right'' leg at $x{\approx}$95\,Mm; cf.\ \fig{F:poynting}) is shown in \fig{F:heating}b for the set of twelve fieldlines. Just as the Ohmic heating at the apex of the fieldlines, this energy input shows a clear (double) peak in time. Comparing the two panels of \fig{F:heating} shows that the energy input at the base of the corona precedes the heating rate at the apex by about 30\,s (the dotted lines in both panels). For a typical Alfv\'en speed of some 500\,km/s this is the time delay expected for the magnetic disturbances traveling up the half loop length of some 15\,Mm from the coronal base.

%------------------------------------------------------------------------------
%------------------------------------------------------------------------------

\newpage

\mysubsection{Hot spot of energy input at the coronal base}%
The discussion above shows clearly the increase and subsequent decrease of the heat input on individual expanding fieldlines. To investigate the cause for this transient heating on a fieldline, we follow the footpoints of the fieldlines at the base of the corona and relate it to the vertical Poynting flux at the base of the corona as defined in \equ{E:Poynting.vertical}.

In \fig{F:poynting} we display the vertical Poynting flux at the time $t{=}$130\,s along with the projection of the twelve selected field lines as defined before in SM~\ref{S:follow}. The pattern of the vertical Poynting flux is relatively stable over the course of more than 10~minutes (cf. the movie attached to \fig{F:poynting}). In particular, the increased Ponyting flux in the middle of the panel remains roughly at the same position forming some sort of \emph{hot spot}. This hot spot at the base of the corona is roughly co-spatial with the increased Poynting flux at the solar surface (the black polygon in \fig{F:poynting} is at the same position as the polygon in \fig{F:photosphere} in the main text showing the Poynting flux at the surface).

While the fieldlines evolve and rise into the atmosphere they move (roughly) horizontally at low heights. Thus at the base of the corona they transverse the hot spot of the Poynting flux. This is evident by inspection of the movie attached to \fig{F:poynting}. Of course, while the fieldline transverses the hot spot, the Poynting flux at the coronal base changes slowly. This is the reason why a dip is seen between two peaks of the Poynting flux  in \fig{F:heating}b. However, it is not the temporal change of the Poynting flux at the base of the corona that is responsible for the increase of the heating. Instead, the main effect for this is the footpoint of the fieldline transversing a hot spot of Poynting flux at the base of the corona.

\formatting{\vspace{0.7ex}}

\begin{figure*}[t]
\noindent
\begin{minipage}[b]{\columnwidth}
\includegraphics[width=\textwidth]{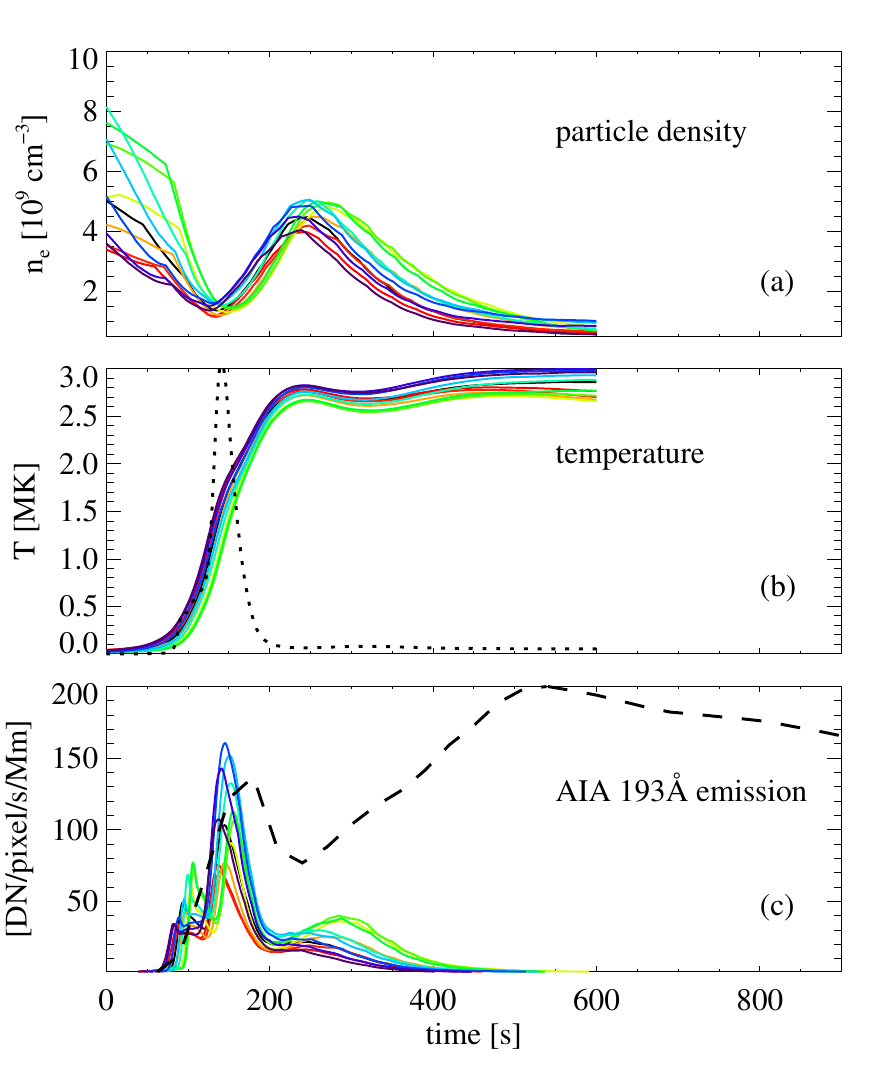}
\end{minipage}
\hfill
\begin{minipage}[b]{\columnwidth}
\caption{
Temporal evolution at the apex of  fieldlines. Panels (a) and (b) show the temporal variation of the temperature and density near the apex of each of the evolving fieldlines. These are the same fieldlines as in \fig{F:heating} with the same color coding. Panel (c) displays the synthesised emission, also at the intersection of the respective fieldline with the vertical midplane. The normalised temperature response curve (for the central fieldline) of the AIA 193\,\AA\ channel dominated by Fe\,{\sc{xii}} is overplotted as a dotted line in panel (b). The emission of the whole EUV loop (integrated over the vertical midplane) is shown as a dashed line in panel (c).
\label{F:thermo}
}
\end{minipage}
\vspace{4ex}
\end{figure*}

\mysubsection{Temperature and density along individual fieldlines}%
While a fieldline is rising upwards through the lower atmosphere (with up to 10\,km/s to 30\,km/s vertically), the density decreases continuously (before time $t{\approx}$130\,s; \fig{F:thermo}a). This is because the rising fieldline is lifting up the cool material and the upwards directed pressure gradient is no longer able to counteract gravity. Thus the plasma drains downwards along the fieldline. In part, the loss of mass for individual fieldlines is also due to numerical imperfections of the simulation: (hyper) diffusion that is needed to smooth out numerical instabilities allows the plasma also to diffuse across fieldlines. In this early phase, when the fieldlines rise through the chromosphere, this can account for up to about half of the mass loss of individual fieldlines. However, this might not be too unrealistic considering that on the real Sun in the cool chromosphere there will be significant cross-field diffusion of mass, because the plasma is only partially ionised there. Once the temperature on the fieldline starts to rise (around $t{\approx}$130\,s) the cross-field mass diffusion no longer plays a role.

%------------------------------------------------------------------------------
%------------------------------------------------------------------------------

After the heating on the fieldline sets in, the temperature will rise (for the set of fieldlines considered here, this happens at $t{\approx}$130\,s; see \fig{F:thermo}b). The draining of the mass from the fieldline and the increase of the heating rate together leads to a very strong increase of the heating \emph{per particle}, which is responsible for the very sharp increase in temperature. Within some 50\,s the peak temperature along the expanding fieldline is rising from basically chromospheric temperature to well above 1\,MK, eventually reaching some 3\,MK. 

In response to the heating of the plasma along the fieldline, heat conduction back to the surface together with the enhanced energy input in the low parts of the atmosphere leads to heating and evaporation of cool plasma into the upper atmosphere. The resulting upflows cause a gradual increase of the density (from time $t{\approx}$130\,s to 250\,s; \fig{F:thermo}a). 

Once the heating on that fieldline ceases (around $t{\approx}$200\,s; \fig{F:heating}), the temperature remains high, because the coronal cooling time is of the order of the better part of an hour. However, the density starts dropping soon after the heating stopped (from time $t{\approx}$250\,s onwards; \fig{F:thermo}a). 
This can be illustrated with the help of long-known equilibrium considerations, even though the variability in the modelled system is more complex. The temperature $T$ and the pressure $p$ (and thus the density $\rho$) are basically set by the heat input $H$; under equilibrium conditions they follow power laws\cite{Rosner+al:1978}, $T{\,\propto\,}H^{2/7}$, $\rho{\,\propto\,}H^{4/7}$, i.e., the density is more sensitive to changes of the heat input than the temperature. Therefore the density adjusts faster to the drop of the heat input after $t{\approx}$200\,s.
(The density drop for each individual fieldline occurs after the temperature passed through the temperature of maximum response of the EUV passband and is thus not of major relevance for the phenomenon described here; see below).

This filling and draining along fieldlines has been described earlier for this 3D model\cite{Chen+al:2014}, and the average rate of change of the mass in the top 20\,Mm of the fieldline is consistent with the (mostly vertical) mass flow across the chromosphere-corona boundary.
In Fig.\,3 of our previous study\cite{Chen+al:2014} the mass exchange is summarised: increased heating causes an evaporative upflow in the bottom part, and later when the fieldline expanded further, the velocity pattern in the upper part reverses and the loop starts draining. (Note that the velocity in that Fig\,3 is along the loop, e.g., red on the left side and blue on the right side does not imply a siphon flow, but evaporation into the corona). In addition to this, the lower transition region is pushed down due to the increase in pressure following the heating in the upper layers, similar to what has been found in quiet Sun network model\cite{Hansteen+al:2010}. Together with the expansion of the fieldlines, this would produce a pattern of net redshifts in the transition region and blueshifts in the hotter regions as found in observations\cite{Peter+Judge:1999}.

This behaviour of the temperature and density along the fieldline is consistent with one-dimensional models of coronal loops with variable prescribed heating rates\cite{Mariska:1987}. However, here we self-consistently describe the heat input along each fieldline in the 3D model as determined by the fieldlines being moved across the hot spot of Poynting flux at the base of the corona.

\formatting{\vspace{2.5ex}}

\mysubsection{EUV emission along individual fieldlines}%
What we see of the corona is neither the temperature nor the density, but the photons that are emitted by the plasma. Thus we synthesise the emission from the model as it would be seen by an EUV instrument. Here we concentrate on the 193\,\AA\ channel of AIA/SDO that images emission from mainly Fe\,{\sc{xii}} forming at about 1.5\,MK. For this we use the same procedures as outlined before\cite{Peter+Bingert:2012,Boerner+al:2012}, implicitly assuming ionisation equilibrium.

The 193\,\AA\ channel has a temperature response function $G(T)$ that peaks sharply at about 1.5\,MK. Consequently, when following an individual fieldline that is heated in time, the contribution to the 193\,\AA\ channel will be significant just during the time the fieldline is at the matching temperature (cf. \fig{F:thermo}b). The actual emission is then given by $n^2\,G(T)$, where $n$ is the (electron) density. To characterize the emission from any given fieldline, we show the emission at the intersection of the fieldline with the midplane used before in \fig{F:thermo}c. This reflects the emission near the apex of the fieldline.

The increase of the coronal emission on an individual fieldline peaks sharply (\fig{F:thermo}c) when the temperature is close to the peak of the contribution function $G(T)$. The timescale for the brightening of an individual fieldline is thus determined by the rise time of the temperature, and is of the order of 50\,s. 
Figure\,\ref{F:thermo} shows that the peak of the coronal emission for each individual fieldline is during the phase of rising density, well before the density drops because the heating for the respective fieldline ceased. Thus the temporal variability of the coronal emission for each individual fieldline is mainly governed by the evolution of the temperature: Each fieldline brightens shortly after it was heated and it temperature rose quickly.

So ultimately each of the expanding fieldlines is brightening according to the time when the footpoint of the fieldline transverses the hot spot of the Poynting flux at the base of the corona causing the enhanced heat input. Consequently the fieldlines lighten up in succession according to their expansion. This is evident from the set of fieldlines shown color coded in \figs{F:heating} and \ref{F:thermo}c for the heat input and emission.

The above discussion concentrates on the results relating to EUV instruments, e.g., the Atmospheric Imaging Assembly (AIA)\cite{Lemen+al:2012}. Currently EUV imaging provides the highest spatial resolution in the corona, significantly higher than X-ray observations, e.g. with the recent XRT instrument\cite{Golub+al:2007}.
However, the response in temperature for X-ray instruments is quite different than EUV instruments. The EUV bands typically show plasma over a temperature range of 0.3 in $\log_{10}T$\,[K] ({\sc{fwhm}} of the response function)\cite{Boerner+al:2012}, i.e.\ a factor of 2. In contrast, the X-ray instruments typically image plasma at higher temperature over wider range of temperature (peak of response function near 8\,MK, covering a factor of 4 in temperature)\cite{Golub+al:2007}. This different response might change the situation quite a bit, in particular because the temperatures in the model loop discussed here reach peak temperatures of about 3\,MK.
Work including the synthesis of X-ray emission to discuss this in more detail is underway.

\begin{figure*}[t]
\noindent
\centerline{\includegraphics[width=0.8\textwidth]{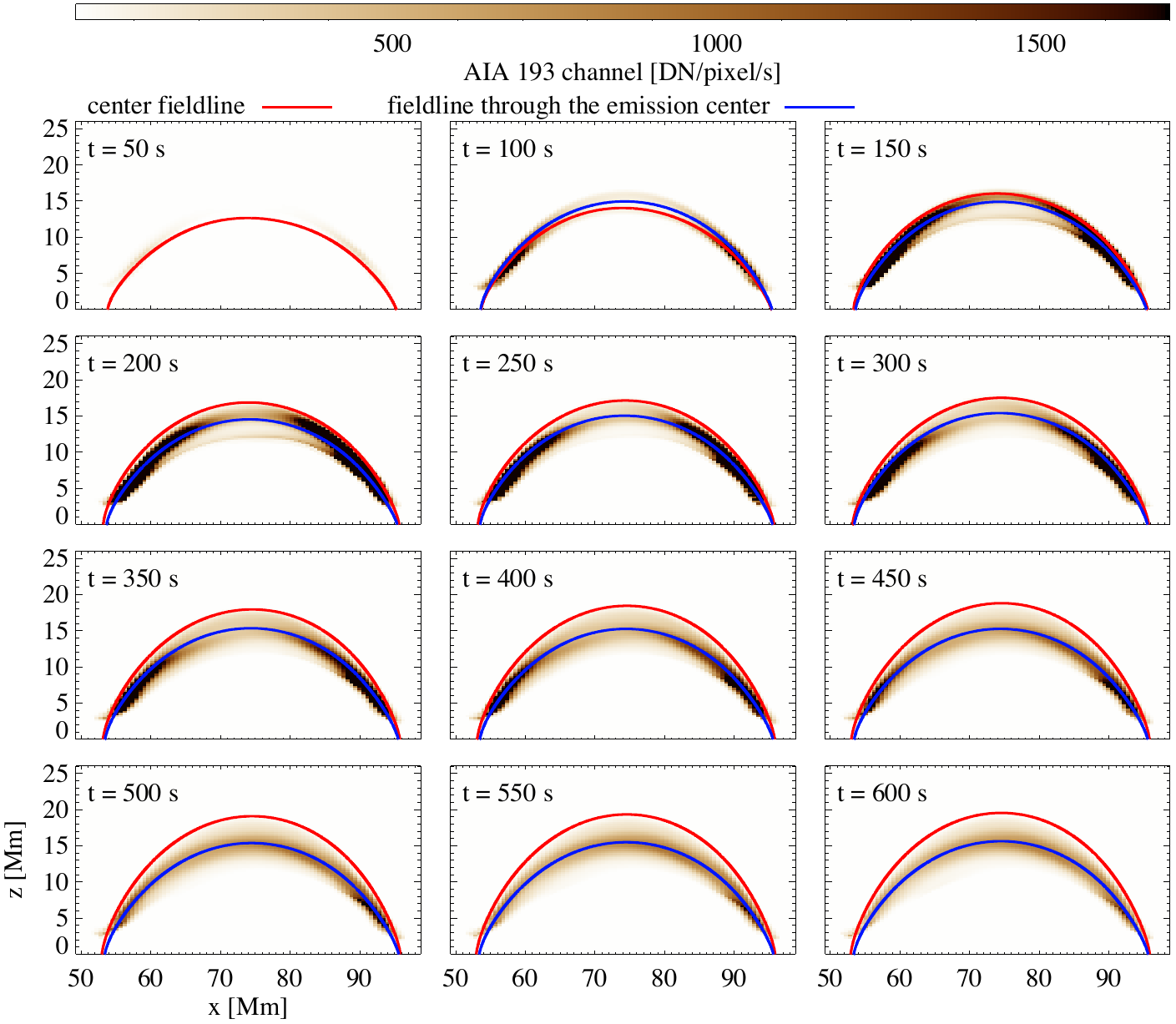}}
\caption{Magnetic fieldline moving through EUV loop. This image sequence shows the evolution of the synthesized EUV emission in the 193\,\AA\ band of AIA dominated by plasma at 1.5\,MK radiating in Fe\,{\sc{xii}}. The red line shows the fieldline in the expanding magnetic field, the blue line the fieldline through the center of the EUV emission pattern at each snapshot. This is similar to panel (a) in \fig{F:side} in the main text.
\newline
The movie showing the full temporal evolution is available online with \fig{F:side} of the main text.
\newline
It is also available at:
%\href[options]{URL}{text}
\href{http://www2.mps.mpg.de/data/outgoing/peter/papers/2015-magnetic-jam/movie-fig-2.mp4}{\color{magenta}http://www2.mps.mpg.de/data/outgoing/peter/papers/2015-magnetic-jam/movie-fig-2.mp4}
\label{F:time.series}
}
\vspace{6.5ex}
\end{figure*}

\mysubsection{Magnetic fieldlines moving\\through stationary EUV loop}%
In \fig{F:time.series} we show the temporal evolution of the resulting EUV loop when seen from the side together with the position of the center fieldline of the magnetic tube selected in SM~\ref{S:follow}. The magnetic tube is constantly moving upwards, from an apex height of ${\approx}$12.5\,Mm at $t{=}$50\,s to ${\approx}$17\,Mm at 200\,s corresponding to a speed of 30\,km/s. The expansion of the fieldlines slows down at greater heights because they are now running into the fieldlines that emerged before. Until the end of the time series shown in \fig{F:time.series} at $t{=}$600\,s the magnetic tube expanded only another 2.5\,Mm, corresponding to an average speed of about 5\,km/s.

%------------------------------------------------------------------------------
%------------------------------------------------------------------------------

While each individual fieldline brightens up for only some 50\,s to 100\,s the successive brightening of the expanding fieldlines causes a comparably stationary bright loop visible in coronal EUV emission (\fig{F:time.series}). In particular the apex height of the center of the EUV loop (blue line in \fig{F:time.series}) varies only slightly between $z{\approx}$14.5\,Mm and ${\approx}$15.5\,Mm, while over the same time the magnetic tube rose by more that 7\,Mm.

The EUV loop formed by the expanding fieldlines shows some variability of its brightness at the apex, but it remains bright for well over 10\,min (dashed line in \fig{F:thermo}c). However, after about $t{\approx}$550\,s the loop starts fading away because the hot spot of the Poynting flux at the base of the corona gets weaker.

The formation of the EUV loop and the apparent motion of the fieldlines though the loop can be summarized as follows. The fieldlines expand during the flux emergence process and start rising. In this process they loose mass because the cool plasma drains along the fieldlines. During the expansion the fieldlines also move horizontally into the forming sunspot. In this process they transverse a hot spot of enhanced Poyntig flux that is formed because of the coalescent flow forming the spots. Thus each individual fieldline gets a burst of heating and when it reaches the response temperature of the EUV channel it brightens up for less than a minute. The expanding fieldlines transverse the hot spot in close succession and thus light one after the other. This then forms an EUV loop being stable for a longer time, basically as long as the hot spot of Poynting flux is sustained by the (horizontal) flows at the base of the corona. Because these fieldlines get bright at a similar geometric height, this creates the illusion that the loop stays at the same position. In reality, the stable-looking EUV loop is formed by a dynamically evolving magnetic field.

\end{multicols}

%%%%%%%%%%%%%%%%%%%%%%%%%%%%%%%%%%%%%%%%%%%%%%%%%%%%%%%%%%%%%%%%%%%%%%%
%%%%%%%%%%%%%%%%%%%%%%%%%%%%%%%%%%%%%%%%%%%%%%%%%%%%%%%%%%%%%%%%%%%%%%%
\end{document}